\input harvmac.tex
%\draftmode
\let\includefigures=\iftrue
\newfam\black
\includefigures
\input epsf
\def\figin{\epsfcheck\figin}\def\figins{\epsfcheck\figins}
\def\epsfcheck{\ifx\epsfbox\UnDeFiNeD
\message{(NO epsf.tex, FIGURES WILL BE IGNORED)}
\gdef\figin##1{\vskip2in}\gdef\figins##1{\hskip.5in}% blank space instead
\else\message{(FIGURES WILL BE INCLUDED)}%
\gdef\figin##1{##1}\gdef\figins##1{##1}\fi}
\def\DefWarn#1{}
\def\figinsert{\goodbreak\midinsert}
\def\ifig#1#2#3{\DefWarn#1\xdef#1{fig.~\the\figno}
\writedef{#1\leftbracket fig.\noexpand~\the\figno}%
\figinsert\figin{\centerline{#3}}\medskip
\centerline{\vbox{\baselineskip12pt
\advance\hsize by -1truein\noindent
\footnotefont{\bf Fig.~\the\figno:} #2}}
\bigskip\endinsert\global\advance\figno by1}
%%%
\else
\def\ifig#1#2#3{\xdef#1{fig.~\the\figno}
\writedef{#1\leftbracket fig.\noexpand~\the\figno}%
%\figinsert\figin{\centerline{#3}}\medskip
\centerline{\vbox{\baselineskip12pt
%\advance\hsize by -1truein\noindent
\footnotefont{\bf Fig.~\the\figno:} #2}}
%\bigskip\endinsert
\global\advance\figno by1}
\fi
%%%%%%%%%%%%%%%%%%%%%%%%%%%%%%%%
\def\co{{\cal O}}
\def\cg{{\cal G}}
\def\cl{{\cal L}}

\def\cn{{\cal N}}
\def\cm{{\cal M}}
\def\ct{{\cal T}}
\def\cp{{\cal P}}
\def\cf{{\cal F}}

\def\p{{\bf{P}}}

%%%%%%%%%%%%%%%%%%%%%%%%%%%%%%%%%%%%%%%%%%
\Title{\vbox{\baselineskip12pt\hbox{hep-th/9811129}
\hbox{IASSNS-HEP-98/99}}}
{\vbox{\vskip-100pt
\hbox{\centerline{Spectral Covers, Charged Matter and}}
\vskip 10pt
\hbox{\centerline{Bundle Cohomology}}}}
\vskip 10pt
\centerline{Duiliu-Emanuel Diaconescu$^\natural$ and 
Gheorghe Ionesei$^\sharp$}
\bigskip
\centerline{\it $^\natural$ School of Natural Sciences}
\centerline{\it Institute for Advanced Study}
\centerline{\it Olden Lane, Princeton, NJ 08540}
\centerline{\tt diacones@sns.ias.edu}
\medskip
\centerline{\it $^\sharp$ Department of Mathematics}
\centerline{\it University ``Alexandru Ioan Cuza''}
\centerline{\it Iasi, 6600 Romania}
\centerline{\tt giones@uaic.ro}
\bigskip
\bigskip
\bigskip
\noindent
We consider four dimensional heterotic compactifications on smooth 
elliptic Calabi-Yau threefolds. Using spectral cover techniques, 
we study bundle cohomology groups corresponding to charged matter 
multiplets. The analysis shows that in generic situations, the 
resulting charged matter spectrum is stable under deformations 
of the vector bundle.

\Date{November 1998}

\newsec{Introduction}

In the recent years, there has been important progress in understanding
the mathematical structure of four dimensional $N=1$ string vacua. 
Most of the new developments have been made possible by the discovery 
of F-theory
\nref\CV{C. Vafa, ``Evidence for F-theory'', Nucl.Phys. {\bf B469}
(1996) 403, hep-th/9602022.}%
\nref\MV{D.R. Morrison and Cumrun Vafa, ``Compactifications of 
F-Theory on Calabi--Yau Threefolds -- I'', Nucl.Phys. {\bf B473}
(1996) 74, hep-th/9602114; ``Compactifications of F-Theory on 
Calabi--Yau Threefolds -- II'', Nucl.Phys. B476 (1996) 437,
hep-th/9603161.}%
\refs{\CV,\MV}
and it's relation to heterotic models. At technical 
level, the duality relates two sets of seemingly very different
geometric data. Namely, the heterotic string vacua are generally 
characterized by a complex holomorphic vector bundle $V$ over a 
$d$-dimensional Calabi-Yau variety $Z$. On the other hand, the 
F-theory models are determined by an elliptically fibered 
$(d+1)$-dimensional Calabi-Yau variety $X$. The problem of mapping the 
geometric moduli of the pair $(V,Z)$ to those of $X$ is very 
complicated. A systematic approach is based on the powerful 
construction of holomorphic bundles on elliptic fibrations presented 
in 
\nref\FMW{R. Friedman, J. Morgan, E. Witten, ``Vector Bundles And F
Theory'', Commun. Math. Phys. {\bf 187} (1997) 679, hep-th/9701162.}%
\nref\BJPS{M. Bershadsky, A. Johansen, T. Pantev, V. Sadov, 
``On Four-Dimensional Compactifications of F-Theory'', 
Nucl. Phys. {\bf B505} (1997) 165, hep-th/9701165.}%
\nref\FMWb{R. Friedman, J.W. Morgan, E. Witten, ``Vector Bundles 
over Elliptic Fibrations'', alg-geom/9709029.}%
\refs{\FMW,\BJPS,\FMWb}. 
Various aspects of this map in different dimensions and 
including certain nonperturbative aspects have been discussed in 
\nref\CL{G. Curio, D. Lust, ``A Class of $N=1$ Dual String Pairs and its
Modular Superpotential'', Int. J. Mod. Phys. {\bf A12} (1997) 5847,
hep-th/9703007.}%
\nref\ACL{B. Andreas, G. Curio, D. Lust, ``$N=1$ Dual String Pairs and 
their Massless Spectra'', Nucl. Phys. {\bf B507} (1997) 175,
hep-th/9705174.}%
\nref\CA{G. Curio, ``Chiral Multiplets in $N=1$ Dual String Pairs'', 
Phys. Lett. {\bf B409} (1997) 185, hep-th/9705197.}%
\nref\AC{B. Andreas, G. Curio, ``Three-Branes and Five-Branes in 
N=1 Dual String Pairs'', Phys. Lett. {\bf B417} (1998) 41,
hep-th/9706093.}%
\nref\PSA{P.S. Aspinwall, ``Aspects of the Hypermultiplet Moduli Space 
in String Duality'', J. High Energy Phys. {\bf 04} (1998) 019,
hep-th/9802194.}%
\nref\CD{G. Curio, R.Y. Donagi, ``Moduli in $N=1$ heterotic/F-theory
duality'', Nucl. Phys. {\bf B518} (1998) 603, hep-th/9801057.}%
\nref\CB{G. Curio, ``Chiral matter and transitions in heterotic 
string models'', hep-th/9803224.}%
\nref\BA{B. Andreas, ``On Vector Bundles and Chiral Matter in $N=1$
Heterotic Compactifications'', hep-th/9802202.}%
\refs{\FMW,\BJPS,\CL,\ACL,\CA,\AC,\PSA,\CD,\CB,\BA}.

Although the string dynamics as well as the duality map are better
understood in eight and six dimensions, phenomenologically interesting
models eventually involve four dimensional compactifications. 
The latter present a series of peculiar aspects, qualitatively 
different from their higher dimensional counterparts. The present work 
is focused on the detailed understanding of one of these aspects, 
namely the charged matter spectrum localized along a codimension 
one locus $\Sigma$ in the base of the elliptic fibration. 
In four dimensional 
models, under suitable conditions, $\Sigma$ is an algebraic curve. 
These charged  
multiplets can be described from dual points of view either 
as zero modes associated to certain bundle cohomology groups 
or as degrees of freedom localized at the intersections of F-theory 
seven-branes. Here we take the first point of view which has the 
advantage of a precise description of the twisting line bundle 
on $\Sigma$. Our analysis 
is focused on charged matter multiplets corresponding to cohomology 
groups of the form $H^1(Z,V)$ which can be localized according to 
\FMW. We prove that the cohomology $H^1(Z,V)$ reduces to the cohomology 
$H^0\left(\Sigma, \cf\right)$ of a twisting line bundle $\cf$ on 
$\Sigma$ and we provide a detailed derivation of $\cf$
using direct image techniques and the Grothendieck-Riemann-Roch 
theorem. Similar issues have been briefly considered in \CB. 
The precise computation 
of the cohomology groups depends on the particular aspects of the 
theory, but it can be performed explicitly in large classes of models. 

The techniques developed here also allow us to address another question 
of interest. In general, we expect a nontrivial variation of the 
cohomology group $H^1\left(Z,V\right)$ under deformation of the 
vector bundle $V$. This is most easily seen by noting that in many 
cases the line bundle $\cf$ corresponds to a special divisor on 
$\Sigma$. Therefore there is a potential variation in the
number of holomorphic sections $h^0\left(\Sigma,\cf\right)$ 
as $\Sigma$ moves in it's linear system on $B$. In physical terms, 
this means that the number of generation anti-generation pairs 
could vary function of the vector bundle moduli. 
We are able to prove that in generic situations i.e. for nonsingular
reduced and irreducible spectral covers, the cohomology is stable 
under deformations of the vector bundle. 
It should be noted that this analysis does not exclude potential 
exceptional behavior which will be investigated elsewhere. 
Also, the approach is purely geometrical,
therefore valid in perturbative string theory. 
Although it would be very interesting to understand how the results
are affected by the full string dynamics, a complete 
treatment seems out of reach at the present stage. 

The paper is structured as follows. Section 2 contains a brief review 
of the spectral cover construction and a detailed treatment of 
localization of cohomology. In section 3, we exploit the results 
of section 2 in order to determine the bundle cohomology and it's 
variation
in a large class of models. 

\newsec{Spectral Covers, Cohomology and Localized Matter}

The purpose of this section is twofold. We start with a brief review of 
the spectral cover construction of heterotic vector bundles
following \FMW,\ mainly aimed at fixing 
notations and conventions. We then provide a more detailed treatment 
of localization of cohomology, clarifying and extending the present 
discussions in the 
literature.

The models considered in this paper are four dimensional 
$E_8\times E_8$ heterotic vacua with compactification data 
$(Z,V_1,V_2)$. Here $\pi:Z\rightarrow B$ 
is a smooth elliptically fibered Calabi-Yau threefold with a 
section $\sigma:B\rightarrow Z$ over a rational surface
$B$. Typically, $B$ is isomorphic to a Hirzebruch  
surface $F_e$ or to a del Pezzo surface $dP_k$. 
$V_1, V_2$ are two holomorphic bundles with 
structure groups $SU(n_{1,2})\subset E_8$ and $c_1(V_{1,2})=0$. 
In the following, $Z$ will be taken to be a smooth Weierstrass model 
\eqn\WA{
zy^2=x^3-axz^2-bz^3}
in $\p\left(\co_B\oplus\cl^2\oplus\cl^3\right)$ with $\cl\simeq
K_B^{-1}$ in order to satisfy the Calabi-Yau condition. $a,b$ are
sections of $\cl^4,\cl^6$.

We concentrate on a single $SU(n)$ bundle $V$ embedded in one $E_8$ 
factor.  Since $c_1(V)=0$, the restriction of $V$ to a generic smooth 
fiber $E_b,\ b\in B$ must be flat, therefore 
$V|_{E_b}$ decomposes as a sum of flat holomorphic
line bundles over $E_b$. These bundles can be described \FMW\ in terms
of spectral data $(C,\cn)$ where $C\subset B$ is a ramified $n$-fold 
cover of $B$ and $\cn$ is a line bundle on $C$. The spectral surface 
$C$ is determined by the equation
\eqn\spcoverA{
a_0+a_2x+a_3y+\ldots+a_nx^{n/2}=0}
if $n$ is even, the last term being $a_nx^{(n-3)/2}y$ for $n$ odd.
The coefficients $a_i$ are sections of $\cm\otimes\cl^{-i}$ with 
no common zeroes,
where $\cm$ is a twisting line bundle pulled back from the base.
Since the base is rational, $\cm$ is uniquely determined 
by the class $\eta=c_1(\cm)$. It follows  that $C$ belongs to the 
linear system $\co_Z(n\sigma)\otimes\pi^*\cm$. 
The bundle $V$ is given by 
\eqn\bundle{
V=p_{Z*}\left(p_C^*\cn\otimes \cp|_{Z\times_B C}\right)}
where $p_Z:Z\times_B C\rightarrow Z$ and 
$p_C:Z\times_B C\rightarrow C$ are the natural projections and 
$\cp=\co_{Z\times_B Z}
\left(\Delta-\sigma_1-\sigma_2\right)\otimes \cl^{-1}$ is 
the Poincar\'e line bundle on $Z\times_B Z$. 
Note that the condition $c_1(V)=0$ is equivalent to 
\eqn\istch{
c_1\left(\cn\right)={1\over 2}\left(K_C-\pi_C^*K_B\right)+\gamma}
where $\gamma$ is a $(1,1)$ class on $C$ in the kernel of\foot{We use 
a distinct notation $\pi_C$ for the restriction of $\pi:Z\rightarrow 
B$ to $C$ in order to distinguish classes pulled back to $C$ from classes
pulled back to the threefold $Z$. This distinction will be important 
further on.}
$\pi_C:C\rightarrow B$ \FMW. The class $\gamma$ can be chosen trivial, 
if the line bundle $K_C-\pi_C^*K_B$ has a square root on $C$. 
Since $K_C=\pi_C^*\eta +n\sigma|_C$, in general this holds if 
\eqn\divcond{
\eta\equiv K_B\ (\hbox{mod}\ 2),\qquad 
n\equiv 0\ (\hbox{mod}\ 2).} 
As explained in \FMW,\ this is the case if the bundle $V$ is symmetric 
i.e. $\tau^*V\simeq V^{\vee}$ where $\tau:Z\rightarrow Z$ is the 
involution of the elliptic fibration. 
If the conditions \divcond\ are not satisfied, $\gamma$ has to be chosen 
of the form \FMW\
\eqn\genconstr{
\gamma=\lambda\left(n\sigma-\pi_C^*\eta-n\pi_C^*K_B\right)}
where $\lambda$ is a suitable half-integral number. 

\subsec{Localized Charged Matter}

In the above models, the charged matter content is determined 
by the bundle cohomology groups of the form 
$H^1(Z,R_i)$ where $R_i$ are associated vector bundles on $Z$ 
defined by the representations $R_i$ of $SU(n)$. 
Throughout this paper we will exclusively consider matter multiplets
associated to $H^1(Z,V)$ (the fundamental representation). The other 
multiplets are also interesting, but the present techniques do not 
provide an equally good control on the relevant cohomology spaces.
Also, the following considerations are restricted to bundles arising 
from nonsingular reduced and irreducible spectral covers $C$. 
The behavior of cohomology in more general situations will be 
studied elsewhere.

As explained in \refs{\FMW,\CB}, the cohomology $H^1\left(Z,V\right)$ 
can be calculated using the Leray spectral sequence. This leads to the 
following exact sequence 
\eqn\extseq{
0\rightarrow H^1\left(B,R^0\pi_*V\right)\rightarrow 
H^1\left(Z,V\right)
\rightarrow H^0\left(B, R^1\pi_*V\right)\rightarrow 
H^2\left(B,R^0\pi_*V\right).}
Therefore the computation of $H^1\left(Z,V\right)$ reduces to the 
computation of the direct images $R^i\pi_*V$, $i=0,1,2$. 
Since this involves certain subtleties, it will be presented in detail 
in the following. 

First, note that since the fibers of $\pi$ are one-dimensional, 
the fiber cohomology groups $H^2\left(Z_b, V_b\right)$, 
$b\in B$ vanish everywhere on $B$. In particular, $R^2\pi_*V\simeq 0$ 
is locally free and the base change 
theorem \ref\RH{R. Hartshorne ``Algebraic Geometry'' 
Springer-Verlag, New York 1993.} (th. III.12.11.),
shows that we have an isomorphism
\eqn\baseA{
\phi^1(b):R^1\pi_*V\otimes k(b)\rightarrow H^1(Z_b,V_b)}
for any $b\in B$. As observed in \FMW,\ the $H^1(Z_b,V_b)$ is 
non-zero if and only if $b$ lies in the codimension one locus 
$\Sigma\subset B$ defined by 
\eqn\codone{
a_n=0.}
Since $a_n$ is a section of the line bundle $\cm\otimes{K_B}^n$, 
this is in fact a divisor in the linear system $|\eta+nK_B|$. 
Throughout this paper, we will assume that the generic divisor 
in $|\eta+nK_B|$ is a nonsingular irreducible holomorphic curve. 
This can be achieved, for example, if $\cm$ is sufficiently ample. 
Therefore, the sheaf $R^1\pi_*V$ is a rank zero coherent sheaf 
supported on $\Sigma$. More precisely, $R^1\pi_*V$ can be 
represented as 
$i_*\cf$ where $i:\Sigma\rightarrow B$ denotes the standard 
embedding and $\cf$ is a rank one coherent sheaf on $\Sigma$. 
Note that $R^1\pi_*V\simeq i_*\cf$ is not a locally free 
sheaf on $B$, therefore the base change theorem shows that the 
natural map
\eqn\baseB{
\phi^0(b):R^0\pi_*V\otimes k(b)\rightarrow 
H^0\left(Z_b, V_b\right)}
cannot be surjective. 
In order to obtain more information, we can use the results of 
\nref\BBHM{C. Bartocci, U. Bruzzo, D. Hern\'andez Ruip\'erez and
J.M. Mu\~noz Porras,''Mirror Symmetry on K3 Surfaces via Fourier-Mukai 
Transform'', Commun. Math. Phys. {\bf 195} (1998) 79, alg-geom/9704023.}%
\nref\HM{D. Hern\'andez Ruip\'erez and J.M. Mu\~noz Porras, ``Structure
of the Moduli of Stable Sheaves on Elliptic Fibrations'', alg-geom/
9809019.}%
\nref\AD{P.S. Aspinwall and R.Y. Donagi, ``The Heterotic String, the 
Tangent Bundle and Derived Categories'', hep-th/9806094.}%
\refs{\BBHM,\HM,\AD} to show that $R^0\pi_*V$ vanishes. More precisely, 
since $\pi:Z\rightarrow B$ is a flat morphism, the natural map 
\baseB\ is injective. We sketch the proof\foot{A more direct argument,
due to P. Aspinwall and E. Witten is based on the definition of the 
sheaf $R^0\pi_*V$. Just note that a section of $R^0\pi_*V$ over a
Zariski open set in $B$ vanishes almost everywhere since $\Sigma$ is 
closed.} which is formally identical to
that of Proposition 2.7. of \BBHM.\ Let $m_b$ denote the ideal 
sheaf of the point $b$ on $B$. Since the morphism $\pi:Z\rightarrow 
B$ is flat, the $\pi^*m_b$ is the ideal sheaf of $Z_b$ on $Z$, therefore
we have an exact sequence
\eqn\exctidA{
0\rightarrow \pi^*m_b\rightarrow \co_Z\rightarrow \co_{Z_b}\rightarrow 0.}
By tensoring with $V$, we obtain 
\eqn\exctidB{
0\rightarrow \pi^*m_b\otimes V\rightarrow V\rightarrow V_b\rightarrow 0}
where $V_b$ is extended by zero to $Z$. This yields by taking direct
images
\eqn\dirimg{
0\rightarrow m_b\otimes R^0\pi_*V\rightarrow R^0\pi_*V\rightarrow
R^0\pi_*V_b\simeq H^0\left(Z_b,V_b\right)}
showing that 
\eqn\inj{
\phi^0(b):R^0\pi_*V_b\otimes k(b)\rightarrow H^0\left(Z_b,V_b\right)}
is injective. 

Since we have also noted that $\phi^0(b)$ cannot be surjective
and $\hbox{dim}H^0\left(Z_b, V_b\right)=1$, it follows that 
$R^0\pi_*V=0$. Taking into account \extseq,\ this shows that 
\eqn\vbcoh{
H^1(Z,V)\simeq H^0\left(B, R^1\pi_*V\right)\simeq 
H^0\left(\Sigma, \cf\right).}
Therefore, in order to complete the computation, we have to determine 
the rank one sheaf $\cf$. A very useful technical result 
\refs{\BBHM,\HM} states that, since the fibers of $\pi:Z\rightarrow B$ 
are one-dimensional, the first image commutes with base change. 
Concretely, given a diagram of the form 
\eqn\diagrA{
\matrix{& Z\times_B B^\prime &{\buildrel p_Z\over\longrightarrow}& Z\cr 
        & & & \cr
        &  {p_{B^\prime}}\biggl \downarrow & & \pi\biggl \downarrow \cr
        & & & \cr
        & B^\prime &{\buildrel u\over\longrightarrow}& B,\cr}}
we have 
\eqn\basechA{
u^*R^1\pi_*V\simeq R^1{p_{B^\prime *}}\left(p_Z^*V\right).}
In general, this relation is not true unless $u:B^\prime\rightarrow B$
is a flat base extension \RH\ (prop. III.9.3.). The fact that it holds 
in this case is a consequence of the base change theorem\foot{
We thank D. Hern\'andez Ruip\'erez for helpful explanations on 
these issues.}.
Applying this result for the diagram 
\eqn\diagrB{
\matrix{& Z\times_B \Sigma &{\buildrel q\over\longrightarrow}& Z\cr 
        & & & \cr
        &  {p_\Sigma}\biggl \downarrow & & \pi\biggl \downarrow \cr
        & & & \cr
        & \Sigma &{\buildrel i\over\longrightarrow}& B,\cr}}
it follows that
\eqn\basechB{
\cf\simeq i^*R^1\pi_*V\simeq R^1p_{\Sigma*}\left(q^*V\right).}
Now consider the diagrams 
\eqn\diagrC{
\matrix{& Z\times_B \Sigma &{\buildrel r\over\longrightarrow}& 
          Z\times_B C \cr 
        & & & \cr
        &  {p_\Sigma}\biggl \downarrow & & {p_C}\biggl 
        \downarrow\cr
        & & & \cr
        & \Sigma &{\buildrel j\over\longrightarrow}& C\cr}
\matrix{&{\buildrel p_Z\over\longrightarrow}& Z\cr
        & & \cr
        & & \pi\biggl \downarrow \cr
        & {\buildrel \pi_C\over\longrightarrow}& B\cr}}
Note that $q=p_Z\circ r$ and $i=\pi_C\circ j$ since $\Sigma=\sigma\cap C$
in $Z$. Applying the same technique, we obtain 
\eqn\basechC{
R^1p_{\Sigma*}\left(q^*V\right)\simeq 
R^1p_{\Sigma*}\left(r^*p_Z^*V\right)\simeq
j^*R^1p_{C*}\left(p_Z^*V\right).}
The last piece of the puzzle is then provided by the relation between 
the vector bundle $V$ on $Z$ and the line bundle $\cn$ on $C$. 
According to \refs{\BBHM,\HM,\AD}, the two objects are related by a 
Fourier-Mukai transform 
\eqn\FM{\eqalign{
&V\simeq R^0p_{Z*}\left(p_C^*\cn\otimes\cp\right)\cr
&\cn\simeq R^1p_{C*}\left(p_Z^*V\otimes\cp^{-1}\otimes 
p_C^*\pi_C^*K_B^{-1}\right).\cr}}
Using again the base change in \diagrC,\ we find that 
\eqn\basechD{
j^*\cn\simeq R^1p_{\Sigma*}
\left(p_Z^*V\otimes p_\Sigma^*j^*\pi_C^*K_B^{-1}\right)}
since the restriction of $\cp^{-1}$ to 
$Z\times_B\Sigma\subset Z\times_B\sigma$ is trivial. 
Therefore, the final result is 
\eqn\basechE{
\cf\simeq R^1p_{\Sigma_*}\left(p_Z^*V\right)\simeq 
j^*\left(\cn\otimes \pi_C^*K_B\right)}which can be rewritten 
\eqn\final{
\cf\simeq j^*\cn\otimes i^*K_B.}
Note that $\cf$ is a line bundle on $\Sigma$ whose 
degree can be easily computed from \istch\ and \genconstr\
\eqn\degf{
\hbox{deg}\cf={1\over 2}\Sigma (\Sigma+K_B)+\lambda 
\left(\Sigma\eta\right).}

The above derivation can also be applied to the dual bundle 
${V^{\vee}}$, leading to an interesting localized interpretation of 
chirality also noted in \CB. 
Relative duality for the finite flat morphism $p_Z:Z\times_B
C\rightarrow Z$ yields
\ref\FMWb{R. Friedman, J.W. Morgan, E. Witten, ``Vector Bundles 
over Elliptic Fibrations'', alg-geom/9709029.} (prop.5.20.)
\eqn\dual{
V^{\vee}\simeq R^0p_{Z*}\left(p_C^*\cn^{-1}\otimes\cp^{-1}\otimes
\omega_{Z\times_B C/Z}\right)}
where $\omega_{Z\times_B C/Z}$ is the relative dualizing sheaf. 
Since $K_Z\simeq \co_Z$, the latter is determined by 
\eqn\relsheaf{
\omega_{Z\times_B C/Z}\simeq \omega_{Z\times_B C}\simeq
p_C^*\left(K_C\otimes\pi_C^*K_B^{-1}\right).}
Therefore we obtain 
\eqn\dualB{
V^{\vee}=R^0p_{Z*}\left(p_C^*\left(\cn^{-1}\otimes K_C
\otimes\pi_C^*K_B^{-1}\right)\otimes \cp^{-1}\right).}
The inverse of \dualB\ is given by 
\eqn\invdual{
\cn^{-1}\otimes K_C\otimes\pi_C^*K_B^{-1}
=R^1p_{C*}\left(p_Z^*V^{\vee}\otimes \cp\otimes p_C^*\pi_C^*
K_B^{-1}\right).}
Repeating the steps \FM-\final,\ we find 
\eqn\dualimg{
R^1\pi_*V^{\vee}\simeq i_*{\cg}}
with 
\eqn\calg{
\cg\simeq j^*\left(\cn^{-1}\otimes K_C\right).}
Comparing \final\ and \calg,\ it follows that 
\eqn\comp{
\cg \simeq K_\Sigma\otimes \cf^{-1}}
where 
\eqn\CYint{
K_\Sigma\simeq j^*K_C\otimes i^*K_B}
since $\Sigma=C\cap \sigma$ in the Calabi-Yau threefold $Z$. 
This yields the promised localization of chirality
\eqn\chiral{\eqalign{
h^1\left(Z,V\right)-h^1\left(Z,V^{\vee}\right)&=
h^0\left(\Sigma,\cf\right)-
h^0\left(\Sigma,K_{\Sigma}\otimes \cf^{-1}\right)\cr
&=\hbox{deg}{\cf}-g\left(\Sigma\right)+1=
-\lambda\left(\Sigma\eta\right)\cr}}
by Riemann-Roch on $\Sigma$. Note that the result agrees with the 
computation of $c_3(V)$ in \CB.

Since the above derivation is quite abstract, it is instructive to 
perform an independent check of the result by applying
Grothendieck-Riemann-Roch theorem for the morphism 
$\pi:Z\rightarrow B$. Although this does not 
completely determine the direct images, it provides significant 
information on their structure. Following 
\ref\F{W. Fulton, ``Intersection Theory'', Springer-Verlag, Berlin 
Heidelberg New-York 1984.}, we have 
\eqn\GRRa{
\hbox{ch}\left(\pi_{!}V\right)\hbox{Td}(B)=
\pi_*\left(\hbox{ch}(V)\hbox{Td}(Z)\right).}
The left hand side of the above equation can be computed 
taking into account that \refs{\FMW,\CB}
\eqn\charcls{\eqalign{
&\hbox{Td}(Z)=1+{1\over 12}\pi^*\left(c_2(B)+11c_1(B)^2\right)
+\sigma\pi^*c_1(B)\cr
&\hbox{ch}(V)=1-\sigma\pi^*\eta-\pi^*\omega
+\lambda\eta(\eta-nc_1(B))w_Z\cr}}
where $\omega$ is a class on the base and $w_Z$ is the fundamental 
class of $Z$. 
Since $\pi_*$ annihilates all classes pulled back from the base, 
we are left with 
\eqn\lhs{\eqalign{
\pi_*\left(\hbox{ch}(V)\hbox{Td}(Z)\right)&=
-\eta-nK_B+\lambda\eta\left(\eta+nK_B\right)w_B\cr
&=-\Sigma+\lambda\left(\eta\Sigma\right)w_B,\cr}}
where $w_B$ is the fundamental class of $B$ and the intersection 
$\left(\eta\Sigma\right)$ is computed on $B$.  
On the other hand, 
\eqn\rhs{
\hbox{ch}\left(\pi_{!}V\right)=\hbox{ch}\left(R^0\pi_*V\right)-
\hbox{ch}\left(R^1\pi_*V\right).}
According to the above analysis, $R^0\pi_*V$ vanishes and 
$R^1\pi_*\simeq i_*\cf$ where $\cf$ is a line bundle on $\Sigma$ 
determined by \final.\
Applying the Grothendieck-Riemann-Roch theorem 
for the immersion $i:\Sigma\rightarrow B$ \F,\ we find 
\eqn\GRRb{
\hbox{ch}\left(R^1\pi_*V\right)\hbox{Td}(B)=
i_*\left(\hbox{ch}(\cf)\hbox{Td}\left(\Sigma\right)\right).}
The right hand side of the above equation can be easily evaluated
obtaining
\eqn\eval{
\Sigma+\left(\hbox{deg}(\cf)-{1\over 2}\hbox{deg}
\left(K_\Sigma\right)\right)w_B.}
Using \istch,\ \final\ and \CYint,\
we are finally left with 
\eqn\fineval{
\Sigma-\lambda\left(\eta\Sigma\right)w_B}
which is in precise agreement with \lhs.\ 

To summarize, we have showed that the computation of $H^1(Z,V)$ reduces
by localization to the computation of the line bundle cohomology 
group $H^0\left(\Sigma, \cf\right)$, where $\Sigma$ is a non-singular 
irreducible curve. As detailed in the next section, this simplification 
allows an explicit evaluation in a large number of cases. Moreover, 
this description can be used to answer another problem of interest, 
namely the dependence of the cohomology groups on the vector bundle
moduli. 

\newsec{Bundle Cohomology and Variation}

As a starting point, 
let us consider the vector bundle moduli in the spectral cover 
realization. According to 
\refs{\FMW,\FMWb}, 
the moduli of the bundle $V$ can be
associated to either deformations of the spectral surface $C$ or 
of the line bundle $\cn\rightarrow C$. If the twisting line bundle 
$\cm$ is sufficiently ample, $H^1\left(C,\co_C\right)=0$ and 
$\cn$ has no deformations. 
\refs{\FMW,\FMWb}. Assuming that this is the case, we first concentrate
on the variation of cohomology under deformations of the 
spectral surface $C$. The line bundle $\cn$ is kept fixed and 
determined by it's first Chern class \istch.\
At a latter stage we will show that, if this condition is relaxed,  
the results are not affected by deformations of $\cn$.

We first consider the $\tau$-invariant case, when 
the conditions \divcond\ are satisfied and $\lambda$ can be set to zero. 
It will be shortly argued that this is in fact the most mathematically 
interesting case. 
Then, the  spectral line bundle $\cn$ is uniquely determined by 
\eqn\cnlb{
\cn=\left(\pi_C^*\cm\otimes\co_Z(n\sigma)|_C
\otimes\pi_C^*K_B^{-1}\right)^{1/2}}
where the square root exists by \divcond.\ 
Note that these conditions actually imply a stronger statement, namely 
that the line bundle $\pi^*(\cm\otimes K_B^{-1})\otimes \co_Z(n\sigma)$ 
admits a square root on $Z$. This leads to a better understanding of the
problem of variation of cohomology as follows. 

The line bundle $\cf$ determined in \basechE\ is isomorphic to the 
restriction of the line bundle 
\eqn\resA{
\left(\pi^*\left(\cm\otimes K_B\right)\otimes \co_Z(\sigma)\right)^{1/2}}
from $Z$ to $\Sigma$. Since $\Sigma$ is included in $\sigma\simeq B$, 
we can first restrict to $\sigma$, obtaining 
\eqn\resB{
\cf\simeq i^*\left(\cm\otimes K_B^{(n+1)}\right)^{1/2}}
where in the right hand side we have now the restriction of a 
{\it fixed} line bundle $\ct=\left(\cm\otimes K_B^{(n+1)}\right)^{1/2}$
on $B$. Therefore the initial problem 
reduces to studying the dependence of the cohomology group 
\eqn\cohA{ 
H^0\left(\Sigma, i^*\left(\cm\otimes K_B^{(n+1)}\right)^{1/2}\right)}
on the moduli of $C$. In generic situations, the spectral surface 
$C$ moves in the linear system $|n\sigma+\pi^*\eta|$ on $Z$
\refs{\BJPS,\FMWb}. 
The number of parameters is $h^0(C,K_C)$ and can be computed using the
Riemann-Roch theorem. Alternatively, the deformations of $C$ can be 
associated to variations of the sections $a_0,a_2\ldots a_n$ 
of $\cm,\cm\otimes\cl^{-2}\ldots\cm\otimes\cl^{-n}$. Since $\Sigma$ 
is defined by $a_n=0$, it is clear that varying $a_0,a_2\ldots a_{n-1}$ 
leaves \cohA\ unchanged. Therefore, the relevant moduli correspond
to deformations of $a_n\in H^0\left(B, \cm\otimes K_B^n\right)$ which 
move $\Sigma$ in the linear system $|\eta+nK_B|$. It follows that 
the variation of the cohomology of $V$ reduces to the variation of 
\cohA\ when $\Sigma$ moves $|\eta+nK_B|$, keeping the line bundle 
$\left(\cm\otimes K_B^{(n+1)}\right)^{1/2}$ fixed on $B$. 
Note that \degf\ and the adjunction formula give
\eqn\degres{
\hbox{deg}\left(\cf\right)=g(\Sigma)-1.}
Therefore the isomorphism class of $\cf$ defines a point in the 
Jacobian variety $J_{g(\Sigma)-1}(\Sigma)$ of degree $g(\Sigma)-1$ 
of $\Sigma$. 
The problem is then equivalent to studying the motion of the 
corresponding point in $J_{g(\Sigma)-1}(\Sigma)$ 
when $\Sigma$ moves in the 
linear system $|\eta+nK_B|$. Potential variations in the cohomology 
can in principle occur since the divisor class associated to 
$\cf$ is special on $\Sigma$
\nref\GH{P. Griffiths, and J. Harris, ``Principles of Algebraic 
Geometry'', Wiley-Interscience, New-York 1978.}%
\nref\G{P. Griffiths, ``An Introduction to The Theory of Special 
Divisors on Algebraic Curves'', Regional Conference Series in Mathematics
no 44, American Mathematical Society, Providence, Rhode Island 1980.}%
\nref\ACGH{E. Arbarello, M. Cornalba, P. Griffiths and J. Harris, 
``Geometry of Algebraic Curves'', vol. 1, Springer-Verlag, 
New-York 1985.}%
\refs{\GH,\G,\ACGH}.
The number of holomorphic sections
can jump if the class of $\cf$ crosses the singular strata 
$W^r_{g(\Sigma)-1}$ of the $\Theta$-divisor. 

If we consider non-invariant cases, when $\lambda\neq 0$, it can be 
easily inferred from \degf\ that the class of $\cf$ is no longer special.
Therefore, the number of holomorphic section is determined by Riemann-Roch
theorem and is independent of deformations of $\Sigma$. 
Since the discussion has been so far rather general, we 
consider next certain concrete situations.

\subsec{Concrete Computation}

Here we consider models with 
the base $B$ isomorphic to a Hirzebruch surface 
$B\simeq F_e$. A smooth Weierstrass model $Z$ requires $e\leq 2$.
In order to avoid certain technical complications, 
we consider $e=2$. The conclusions are also valid for $e=0,1$ but 
the analysis has to be modified on a case by case basis.
Let $C_0$, $f$ denote the standard generators of the Picard group 
of $B$, satisfying $C_0^2=-2$, $f^2=0$, $C_0\cdot f=1$. Note that the 
canonical class of $B$ is given by 
\eqn\canbase{
K_B=-2C_0-4f.}
We pick the class of $\Sigma$ to be of the form
\eqn\classA{
\left[\Sigma\right]=aC_0+bf,\qquad a,b\in Z_{+}.}
According to \RH\ (cor. V.2.18.), the linear system $aC_0+bf$ contains 
an irreducible nonsingular curve different from $C_0,\ f$ if and only if 
$a>0$ and $b\geq 2a$. 
The conditions \divcond\ are satisfied if
\eqn\divcondB{
a\equiv 0\ (\hbox{mod}\ 2),\qquad b\equiv 0\ (\hbox{mod}\ 2)}
and $n$ is even. Note that this implies $a\geq 2$, $b\geq 4$. 

The class of the fixed line bundle $\ct$ in \resB\ 
is of the form
\eqn\classB{
\left[\ct\right]={1\over 2}\left((a-2)C_0+(b-4)f\right).}
Taking into account the above restrictions on $a,b$, the 
generic divisor in the above linear system is an irreducible nonsingular 
curve $\Gamma$ if $a>2$. 
Hence, the  line bundle $\ct$ can be taken of the form 
$\co_B\left(\Gamma\right)$ for fixed $\Gamma$. 
If $a=2$, it follows from \classB\ that $\Gamma$ can be taken a 
collection of disjoint $P^1$ fibers of $B$. 
Our problem is then to 
compute $H^0\left(B,\co_{\Sigma}(\Gamma)\right)$ and to understand it's
eventual variation. Consider the exact sequence\foot{We thank I. 
Dolgachev for explaining this line of argument to us.}
\eqn\exctA{
0\rightarrow \co_B\left(\Gamma-\Sigma\right)\rightarrow 
\co_B\left(\Gamma\right)\rightarrow \co_\Sigma\left(\Gamma\right)
\rightarrow 0.}
The associated long exact cohomology sequence yields 
\eqn\exctB{\eqalign{
0&\rightarrow H^0\left(B, \co_B\left(\Gamma\right)\right)
\rightarrow H^0\left(\Sigma, \co_\Sigma\left(\Gamma\right)\right)
\rightarrow H^1\left(B, \co_B\left(\Gamma-\Sigma\right)\right)\cr
&\rightarrow H^1\left(B, \co_B\left(\Gamma\right)\right)
\rightarrow\ldots\cr}} 
Now consider the exact sequence 
\eqn\exctC{
0\rightarrow \co_B\rightarrow \co_B\left(\Gamma\right)\rightarrow
\co_\Gamma\left(\Gamma\right)\rightarrow 0.}
The last term is the normal bundle of $\Gamma$, $N_{\Gamma/B}$, 
extended by zero to $B$. Since $H^i\left(B,\co_B\right)=0$ for $i>0$,
the long exact cohomology sequence yields
\eqn\exctD{
0\rightarrow H^1\left(B, \co_B\left(\Gamma\right)\right)
\rightarrow H^1\left(\Gamma, N_{\Gamma/B}\right)\rightarrow 0.}
Next, we have
\eqn\normdeg{\eqalign{
&\hbox{deg}\left(N_{\Gamma/B}\right)=\Gamma^2\cr
&2g\left(\Gamma\right)-2=\Gamma\left(\Gamma+K_B\right).\cr}}
A direct computation shows that if $a>2$, $\Gamma\cdot K_B<0$, hence 
\eqn\nonsp{
2g\left(\Gamma\right)-2<\hbox{deg}\left(N_{\Gamma/B}\right)}
and the divisor $N_{\Gamma/B}$ is not special on $\Gamma$. 
Therefore we obtain
\eqn\vanA{
H^1\left(B, \co_B\left(\Gamma\right)\right)\simeq 
H^1\left(\Gamma, N_{\Gamma/B}\right)\simeq 0.}
If $a=2$, $\Gamma$ is a sum of disjoint rational fibers whose normal 
bundles are trivial, therefore \vanA\ is still valid. 
Moreover, since 
\eqn\id{
\left[\Gamma\right]={1\over 2}\left(\left[\Sigma\right]+K_B\right),}
Kodaira-Serre duality on $B$ shows that 
\eqn\vanB{
H^1\left(B,\co_B\left(\Gamma-\Sigma\right)\right)\simeq
H^1\left(B,\co_B\left(\Gamma\right)\right)^{\vee}\simeq 0.}
Consequently, \exctB\ reduces to 
\eqn\exctE{
0\rightarrow H^0\left(B, \co_B\left(\Gamma\right)\right)
\rightarrow H^0\left(\Sigma, \co_\Sigma\left(\Gamma\right)\right)
\rightarrow 0.}
This shows that the cohomology group 
$H^0\left(\Sigma, \co_\Sigma\left(\Gamma\right)\right)$ does not vary as 
$\Sigma$ moves in it's linear system. Furthermore, it can be explicitly 
computed as the cohomology of a fixed line bundle on the base,
which follows from Riemann-Roch theorem. To this end, 
note that
\eqn\vanC{
H^2\left(B,\co_B\left(\Gamma\right)\right)\simeq 
H^0\left(B, \co_B\left(K_B-\Gamma\right)\right)\simeq 0.}
Since we have also proved that the first cohomology group vanishes, 
the Riemann-Roch formula yields
\eqn\RR{
h^0\left(\co_B\left(\Gamma\right)\right)={1\over 2}\Gamma
\left(\Gamma-K_B\right)+1={1\over 4}a(b-a).}

Collecting all the results of this section, we have shown that in a 
large class of models, the cohomology groups $H^1(Z,V)$ can be precisely
computed function of spectral data and they are stable under 
deformations of the vector bundle. Similar methods can be 
applied in other cases as well, leading to similar conclusions (for 
example when the base $B$ is a rational elliptic surface). Inspecting 
the above chain of arguments suggests that this is the generic behavior
in elliptic models over rational surfaces when the class $\eta$ is 
sufficiently ample.

As mentioned in the beginning of the section, the analysis has been 
so far restricted to 
spectral surfaces $C$ with $H^1\left(C,\co_C\right)=0$. 
In the following, we show that the conclusions are still valid when 
this condition is relaxed. 

\subsec{Nontrivial $H^1\left(C,\co_C\right)$}

According to \FMWb\ (lemma 5.16), $H^1\left(C,\co_C\right)$ 
can be computed starting from the exact sequence 
\eqn\exctF{
0\rightarrow \co_Z(-n\sigma)\otimes \pi^*\cm^{-1}
\rightarrow \co_Z\rightarrow \co_C\rightarrow 0}
which yields 
\eqn\exctG{\eqalign{
\ldots&\rightarrow H^1\left(Z,\co_Z\right)\rightarrow
H^1\left(C,\co_C\right)\rightarrow 
H^2\left(Z, \co_Z(-n\sigma)\otimes \pi^*\cm^{-1}\right)\cr
&\rightarrow H^2\left(Z, \co_Z\right)\rightarrow\ldots\cr}}
Since in our case $Z$ is a Calabi-Yau variety, 
$H^1\left(Z,\co_Z\right)\simeq H^2\left(Z, \co_Z\right)\simeq 0$, 
therefore we obtain an isomorphism 
\eqn\isom{
H^1\left(C,\co_C\right)\simeq 
H^2\left(Z, \co_Z(-n\sigma)\otimes \pi^*\cm^{-1}\right).}
The right hand side can be evaluated using the Leray spectral sequence
\eqn\evalrhs{\eqalign{
H^1\left(C,\co_C\right)\simeq 
&H^1\left(B, \cl^{-1}\otimes\cm^{-1}\right)\oplus 
H^1\left(B, \cl\otimes\cm^{-1}\right)\oplus\ldots\cr
&\oplus H^1\left(B, \cl^{n-1}\otimes \cm^{-1}\right).\cr}}
Now consider the exact sequence 
\eqn\exctH{
0\rightarrow\co_C(-\Sigma)\rightarrow 
\co_C\rightarrow \co_\Sigma\rightarrow 0.}
The associated long exact cohomology sequence reads 
\eqn\exctI{\eqalign{
0&\rightarrow H^0\left(C, \co_C\right){\buildrel r\over\rightarrow}
H^0\left(\Sigma, \co_\Sigma\right)\rightarrow 
H^1\left(C,\co_C(-\Sigma)\right)\cr
&\rightarrow H^1\left(C, \co_C\right){\buildrel f\over\rightarrow}
H^1\left(\Sigma, \co_\Sigma\right)\rightarrow\ldots\cr}} 
The map $r$ is clearly surjective since any constant function on 
$\Sigma$ can be extended on $C$, therefore the sequence truncates as
\eqn\exctJ{
0\rightarrow H^1\left(C,\co_C(-\Sigma)\right)
\rightarrow H^1\left(C, \co_C\right){\buildrel f\over\rightarrow}
H^1\left(\Sigma, \co_\Sigma\right)\rightarrow\ldots}
By tensoring the exact sequence \exctF\ by $\co_Z(-\sigma)$, 
we obtain
\eqn\exctK{
0\rightarrow \co_Z(-(n+1)\sigma)\otimes \pi^*\cm^{-1}
\rightarrow \co_Z(-\sigma)\rightarrow \co_C(-\Sigma)\rightarrow 0.}
Note that we can use the Calabi-Yau property of $Z$ and the 
rationality of $\sigma$ to show that
$H^1\left(Z, \co_Z(-\sigma)\right)\simeq 0$ and 
$H^2\left(Z, \co_Z(-\sigma)\right)\simeq 0$. This follows from the 
long cohomology exact sequence associated to 
\eqn\exctL{
0\rightarrow \co_Z(-\sigma)
\rightarrow \co_Z\rightarrow \co_\sigma\rightarrow 0.}
Therefore we can repeat 
the steps \exctF-\evalrhs\ to derive
\eqn\derv{
H^1\left(C,\co_C(-\Sigma)\right)\simeq H^1\left(C,\co_C\right)
\oplus H^1\left(B, \cl^n\otimes \cm^{-1}\right).}
However, the line bundle 
$\cl^n\otimes \cm^{-1}$ is isomorphic to $\co_B(-\Sigma)$
and $H^1\left(B, \co_B(-\Sigma)\right)\simeq 0$ by an exact sequence 
argument similar to those presented so far. 
Therefore, the complex vector spaces $H^1\left(C,\co_C(-\Sigma)\right)$
and $H^1\left(C,\co_C\right)$ are isomorphic, implying that the 
linear map $f$ maps $H^1\left(C,\co_C\right)$ to zero in \exctJ. 
This shows that deformations of the line bundle $\cn$ on $C$ are 
mapped to trivial deformations of the restriction $\cn|\Sigma$.
Hence they have no effect on the cohomology, as claimed before. 
\centerline{\bf Acknowledgments}
We are very grateful to Paul Aspinwall, Gottfried Curio, Igor Dolgachev, 
Daniel Hern\'andez Ruip\'erez, Liviu Nicolaescu, Govindan Rajesh
and Edward Witten for very helpful discussions and suggestions.
The work of D.-E. D. has been supproted by grant 
$\sharp$DE-FG02-90ER40542.

\listrefs
\end